\documentclass[
aps,%
12pt,%
final,%
notitlepage,%
oneside,%
onecolumn,%
nobibnotes,%
nofootinbib,%
superscriptaddress,%
noshowpacs,%
centertags
]{revtex4} 

\usepackage{amsmath}
\usepackage{cmap}
\usepackage[usenames]{color}
\usepackage[utf8]{inputenc} 
\usepackage[T2A]{fontenc} 
\usepackage[english,russian]{babel} 
\usepackage{graphicx} 
\usepackage{pdflscape} 
\usepackage[normalem]{ulem}
\begin{document}
\gdef\capnames{
    \gdef\figurename{Fig.}
    }
\renewcommand{\figurename}{Fig.}
\def\figurename{Fig.}
\def\figuresname{Fig.}

\noindent {\it Published in Astronomy Reports, 2019, vol. 63, № 8, pp. 633-641}
\bigskip\bigskip \smallskip\hrule
\noindent {\it}

\title{The Photolysis of Aromatic Hydrocarbons Adsorbed on the Surfaces of Cosmic Dust Grains}

\author{\bf  \quad \firstname{M.~S.}~\surname{Murga}}
\email{murga@inasan.ru}
\affiliation{Institute of Astronomy, Russian Academy of Sciences, Moscow, 119017 Russia}
\affiliation{Lomonosov Moscow State University, Moscow, Russia}

\author{\bf \firstname{V.~N.}~\surname{Varakin}}
\affiliation{Lomonosov Moscow State University, Moscow, Russia}

\author{\bf\firstname{A.~V.}~\surname{Stolyarov}}
\affiliation{Lomonosov Moscow State University, Moscow, Russia}

\author{\bf\firstname{D.~S.}~\surname{Wiebe}}
\affiliation{Institute of Astronomy, Russian Academy of Sciences, Moscow, 119017 Russia}

\begin{abstract}
\centerline{\footnotesize Received October 30, 2018; revised November 27, 2018; accepted November 27, 2018}\bigskip\bigskip\bigskip

Abstract — The work is devoted to the adaptation of the results of laboratory studies of the laser-induced dissociation of molecules of benzene adsorbed on a quartz substrate to the conditions of the interstellar medium. Adsorption was performed under conditions of low temperature and deep vacuum. The difference between the photolysis of adsorbed molecules and molecules in the gas phase is identified. Significance of process of photolytic desorption in the interstellar conditions is analyzed, in particular, in the conditions of photodissociation regions. It is shown that the efficiency and dissociation channels of photolysis of adsorbed and gas phase benzene differ substantially. It is concluded that the photolysis of aromatic hydrocarbons adsorbed on the interstellar dust grains contributes a negligible fraction to the abundance of small hydrocarbons in the interstellar medium.

DOI: 10.1134/S1063772919080043

\end{abstract}

\maketitle

\section{INTRODUCTION}
Spectral observations of nebulae in the infrared (IR) led to the discovery of emission bands at wavelengths of 3– 20~$\mu$m. It was later elucidated that these are generated not only in nebulae, but also in the more diffuse interstellar medium (ISM)~\cite{gillett73, sellgren83, aitken81}. It is now widely believed that the appearance of these bands (which remain unidentified to this day) is due to vibrations in molecules of large polycyclic aromatic hydrocarbons (PAHs)~\cite{leger84,allamandola99}. In the rarified medium of diffuse clouds, isolated PAH molecules are located exclusively in the gas phase, while, in dense interstellar clouds, they can settle onto the surfaces of large cosmic dust grains together with other chemical compounds, forming organic mantles~\cite{sellgren95, allamandola11}.

The brightest PAH emission is observed in so-called photodissociation regions (PDRs), which are interface regions at the peripheries of dense molecular clouds illuminated by the ultraviolet radiation of nearby massive stars. In such regions, in addition to chemically stable polyatomic PAHs, active hydrocarbon  radicals (possessing  unpaired electrons) are observed, such as C$_2$H, C$_3$H$_2$, and so forth~\citep{pety05, guzman15}. One source of these chemically active fragments with small numbers of atoms is the destruction (dissociation or dissociative ionization) of PAH molecules under the action of intense ultraviolet (UV) and vacuum UV (VUV) radiation.

The photostability of aromatic compounds, including PAHs, in the gas phase has been considered in many studies~\cite{jochims94, allain96, visser07, lepage01, murga16a}. It was shown that isolated molecules with fewer than 40–50 carbon atoms are unstable under the conditions of the ISM, while larger PAHs predominantly lose H atoms or are ionized as a result of interactions with UV photons~\cite{zhen15}. In PDRs, PAH molecules can be distributed between the gas phase and organic mantles of dust grains. In this case, powerful UV and VUV radiation from massive stars acts not only on gas-phase PAH molecules, but also on PAH molecules adsorbed on the surfaces of cosmic dust grains; however, the results of this latter interaction are less clear. 

The appreciable abundances of complex organic molecules observed in regions of star formation, which cannot be explained by gas-phase reactions, led to the suggestion that their presence is due to processes occurring on grain surfaces, including photochemical (photo-induced) processes. Laboratory investigations of photochemical reactions in astrophysical ices (e.g., \cite{gerakines96, cottin03, oberg10}) show that, as a rule, the cross sections for photodissociation of adsorbed molecules are appreciably lower than the corresponding cross sections for the same molecules in the gas phase\cite{oberg16}. To a significant extent, this could be related to the fact that fragments of dissociated molecules in a grain mantle can efficiently recombine, reforming the original molecule.

In our current study, we have considered another factor influencing the photostability of a number of adsorbed molecules. A series of laboratory experiments on the photostability of adsorbed molecules of the simplest aromatic hydrocarbons, such as benzene~\cite{varakin18}, chlorobenzene~\cite{varakin19}, and toluene~\cite{varakin16} have been carried out in the Chemistry Department of Moscow State University using powerful UV laser radiation. It was established in these experiments that adsorbed molecules are much more stable than their gas-phase analogs, due to the transfer of the excess internal energy of the photoexcited molecule to the substrate via rapid intramolecular relaxation. In addition, it was concluded based on high-sensitivity mass-spectrometer measurements that physically adsorbed aromatic hydrocarbons can efficiently photodissociate when the light flux is substantially enhanced, becoming an additional source of active molecular components (radicals and ions) in the gas phase.

The described properties of photolytic processes (photo-induced dissociation) with the participation of adsorbed molecules are not taken into account in modern astrochemical models. As a rule, it is assumed that photolysis occurs at the same rates and with the same products as in the gas phase. Obviously, this makes the results of astrochemical modeling less trustworthy. Refining the models requires laboratory investigations and the adaptation of the results to the conditions of the ISM. The main aim of our current study is the adaptation of the results of laboratory experiments on laser-induced dissociation of adsorbed aromatic hydrocarbons to realistic conditions in the ISM, specifically the conditions in PDRs.

\section{LASER-INDUCED DISSOCIATION OF ADSORBED MOLECULES}

We based our study on the results of experiments on laser-induced dissociation of adsorbed molecules carried out in the Laser Chemistry Group of the Chemistry Department of Moscow State University~\cite{varakin18}. A single-layer coating of benzene molecules on a surface of fused quartz cooled with liquid nitrogen to a temperature of $\sim100$~K was deposited at a pressure of $2\times 10^{-5}$~Pa over a time of 10~s. The substrate with the physically adsorbed benzene molecules was irradiated by a pulse of radiation from a KrF excimer laser (wavelength $\lambda=248$~nm, pulse duration $\Delta t_{\rm l}=20$~ns) with an energy density $F_{\rm l}$ that was varied from 20 to 120~mJ~cm$^{-2}$. This energy was sufficient to give rise to both dissociation of the benzene molecule and desorption of the dissociation products. Note that the use of laser radiation is not of fundamental importance. Under the conditions of the experiment, photodissociation is due to absorption of individual photons, and the coherence of the radiation plays no role. Laser radiation is also used in other experiments having an astrochemical context (see, e.g., \cite{zhen14a, zhen14b}).

The identification and measurement of the relative output of fragments desorbed by photodissociation were realized using a Riber QMM-17 quadrupole mass spectrometer located 38 mm from the substrate surface. In the mass spectrometer, fragments were ionized by electrons with energies of 70~eV, after which their mass distribution was analyzed.

Mass-spectrometer signals from atomic hydrogen H, the phenyl radical C$_6$H$_5$, acetylene C$_2$H$_2$, and ethylene C$_2$H$_4$ were detected, suggesting that H atoms were detached from the periphery of benzene molecules, and that some benzene rings were destroyed. The dependence of the mass-spectrometer signals for benzene fragments on the energy density of the laser radiation was measured, and found to be described well by a power law with index $n\approx3$. The proposed mechanism for dissociation of adsorbed molecules included two-photon excitation of the molecules to a high vibronic states located in the quasicontinuum; substantial lengthening of the broken molecular bond (and partial vibrational relaxation of the molecule); and transition of the molecule to a dissociative electronic state outside the quasicontinuum due to the stimulated emission of a photon. This last process occurred when, during its vibrational dynamics, a molecule ended up in a state whose energy exceeded the energy at some point in the dissociative potential curve precisely by an amount equal to the energy of a laser photon. In this case, the energy at this point differs little from the dissociation energy of the bond. As a result, the fragments had low kinetic energies (tens of meV). This mechanism can explain the dissociation of intramolecular bonds with energies lower than the photon energy. This dissociation mechanism is based on the excitation of molecules to a dissociative state with long lengths of the intramolecular bonds. In contrast to gas-phase molecules, where the fragmentation of the molecules can occur through pre-dissociation, or ``hot'', vibrational states, rapid vibrational relaxation with the resonance transmission energy to the substrate hinders the realization of such mechanisms in adsorbed molecules. Their direct photodissociation requires excitation to repulsive states. The presence of minima in the potential-energy curves leads to the excited molecules falling into these minima in the course of their vibrational relaxation. The reason for the absence of states with monotonically decreasing potential energy (purely repulsive states) in the region of the quasicontinuum is the prohibition against the intersection of states with the same symmetry. Direct photoexcitation of repulsive states outside the quasicontinuum from the minimum of the ground state is often inefficient due to the symmetry restriction of the dissociative state.

In the case of laser irradiation, physically adsorbed benzene molecules form three times as many hydrogen atoms as acetylene molecules.	With an energy density for the KrF laser emission of 40~mJ~cm$^{-2}$, the hydrogen-atom signal $S_{\rm H}$ was $2.7\times10^{12}$~atoms~cm$^{-2}$, while the corresponding signal for acetylene molecules was $S_{\rm C_2H_2} =0.9\times10^{12}$~molecules~cm$^{-2}$. With a density of adsorbed benzene molecules $S_0 =1.3\times10^{14}$~cm$^{-2}$, this means that roughly 0.7\% of benzene molecules dissociated with the formation of acetylene, and 2.1\% dissociated with the formation of atomic hydrogen and the phenyl radical. No hydrogen molecules were detected. Their formation preceeds the isomerization of benzene --- a grouping of two hydrogen atoms and one of the carbon atoms~\cite{Kislov04} --- which is hindered in an adsorbed state.

The physical desorption of benzene molecules occurred as a result of heating of the fused quartz, with the absorption of radiation by surface defects. These electron defects, and also microfractures on the surface, form due to the action of the powerful laser radiation. Similar properties may also characterize the surfaces of grains in the ISM, where defects can both be present intrinsically and be formed by the action of high-energy particles and radiation.

\section{CHARACTERISTICS
OF PHOTODISSOCIATION OF AROMATIC HYDROCARBONS ADSORBED ON COSMIC GRAIN SURFACES}

It is usually supposed in astrochemistry that surface reactions proceed in the same way as gas-phase reactions~\cite{Esplugues16}. However, in the dissociation of gas-phase PAHs, the main products are hydrogen atoms and molecules, as well as acetylene~\cite{jochims94}, with the output of molecular hydrogen higher than the output of acetylene. In the experiment with adsorbed benzene molecules, the main products observed were atomic hydrogen and acetylene. Other molecular compounds were also detected, but no molecular hydrogen was observed. This indicates that the photodissociation of gas-phase and adsorbed molecules proceeds not only at different rates, but also along different channels, making it incorrect to directly apply data on the mechanisms and rates of gas-phase reactions to surface chemistry.

The potential-energy surface for the ground (minimum-energy) electronic state of physically adsorbed molecules apparently essentially coincides with the analogous state for isolated molecules in the gas phase, while highly excited electronic states of adsorbed molecules and molecular clusters (in a multi-layer coating) can, according to quantum chemical electron-structure, calculations undergo substantial changes, leading to the formation of so-called supramolecular complexes, in particular, molecular dimers. These complexes are characterized by:
\begin{itemize}
\item an appreciable increase in the density of electron-vibrational states (the quasicontinuum) and a substantial acceleration in intramolecular relaxation (due to the removal of constraints on the selection rules and the formation of numerous energetic resonances with vibrations of the crystal grid of the substrate);
\item weakening of the energies of interatomic bonds, which may lower the threshold for dissociation and ionization, and may sometimes lead to increased probabilities of electronic transitions;
\item heating of a very thin surface layer of the substrate by radiation and the thermal desorption of molecules.
\end{itemize}

Laboratory experiments also indicate very low kinetic energies for the products of photolytic dissociation, compared to those for gas-phase dissociation; this inevitably leads to chemical inertness of the neutral particles formed due to their inability to overcome the corresponding activation barriers. We should note that the enhanced reactivity of the products of gas-phase photolytic dissociation is likewise not taken into account in astrochemical models.

When adapting laboratory studies to the conditions in the ISM, it is important to take into account the thickness of the deposited molecular layer, since the photodissociation parameters for single-layer and multi-layer coatings differ appreciably. A multi-layer coating is unstable against UV radiation, since photoexcitation of the molecules lengthens their intramolecular bonds and breaks weak, intermolecular bonds with unexcited molecules. This gives rise to
desorption of individual molecules and/or molecular clusters, while desorption of molecules from a single- layer coating occurs in accordance with the theoretical mechanism. The subsequent photodissociation of the desorbed particles is analogous to the gas-phase process~\cite{varakin16}. Under the conditions of the ISM, multi-layer coatings consisting of many molecular components predominate. The icy mantles of dust grains contain water, methanol, ammonia, carbonic gas, and other compounds. Therefore, enhancing the relevance of laboratory experiments for these conditions requires determining the compositions of the mantles of interstellar grains as completely as possible, in order to model the photodissociation of aromatic molecules in a realistic chemical environment.

The aromatic compounds in the ISM are probably appreciably larger then benzene. Thus far, no specific PAH molecule has been unambiguously identified in the ISM, but observed bands are very close to PAH systems with 40–100 carbon atoms and with compact structures, and a set of 40–50 such PAHs can explain the observed IR spectra~\cite{andrews15}. Thus, proposed interstellar PAHs are substantially larger and more stable than benzene, and they may not dissociate with the same efficiency as benzene in the adsorbed state. However, all this refers mainly to PAHs in the diffuse ISM. PAHs formed in molecular clouds~\cite{sandstrom10} and contained in the mantles of dust grains are small in size, and the results of the laboratory experiment described above are relevant for these PAHs (if they fall within the zone of influence of a source of UV radiation).

\section{ESTIMATES OF THE EFFICIENCY OF PHOTODISSOCIATION IN THE ISM}

The interstellar radiation field (ISRF) differs substantially from the radiation of the KrF excimer laser. On the one hand, the ISRF encompasses a wide range of wavelengths, so that there are a large number
of combinations of photons that are able to excite an adsorbed molecule in the quasicontinuum and bring about a dissociative state. On the other hand, the intensity of the ISRF is much lower than the intensity of the laser.

To investigate the applicability of the experimental results to the ISM, we compared the parameters of the ISRF with those of the KrF laser that was used. We took the parameters of the ISRF to be those of the mean radiation field in the solar vicinity described in \cite{mmp83}~(MMP83). The spectral flux of the radiation at a wavelength of 248 nm is $\approx1.3\times 10^{13}$~eV~s$^{-1}$~cm$^{-3}$. Given that the duration of the laser pulses was 20~ns and the width of the line spectrum generated by the laser was 3~nm, the energy density of an ISRF with these temporal and frequency parameters
will be $F_{\rm ISRF}\approx 2\times 10^{-2}$~eV~cm$^{-2}$. The energy density of the laser $F_{\rm l}$ at which the absolute measurements of the fragment signals were conducted is $\approx7\times 10^{17}$~eV~cm$^{-2}$. Thus, the energy density of the laser exceeds the corresponding value for the ISM by more than 19 orders of magnitude. This means that the expected signal for three-photon dissociation is 58 orders of magnitude lower in the ISM than in the laboratory. However, this estimate was obtained without taking into account the presence of photons of other energies in the ISM. A possible simplying assumption is that all photons in the ISM have energies of 5~eV and can lead to the dissociation of adsorbed molecules, analogous to the corresponding processes in the experiment. For this fictitious energy density, we find an energy density of $\approx150$~eV~cm$^{-2}$. In this case, the difference between this value and the laser energy density is only 15 orders of magnitude and the expected signal from fragments under the conditions of the ISM is a factor of $10^{45}$ lower than in the laboratory --- a very low value, as previously. It may seem excessive to carry out an extrapolation under these circumstances, but we are interested in order of magnitude estimates indicating the very low efficiency of this process under the conditions of the ISM. Such an extrapolation is admissible for obtaining such qualitative results.

Dissociation will be more efficient if the excitation of the adsorbed benzene molecules in the vibronic quasicontinuum occurs not as a result of a two-photon process involving UV photons, but instead via a resonance absorption of a single VUV photon. The dissociation of adsorbed molecules by the KrF laser described in \cite{varakin18} supposes the absorption of two photons to bring a molecule to the quasicontinuum and the stimulated emission of a third photon to bring about a transition to a dissociative state. When the molecule makes a transition between levels $i-1$ and $i$ after the absorption of the $i$th photon ($i=1,2,3$), the kinetic equation for populating the level $N_i$ can be written

\begin{equation}
\frac{dN_i}{dt} = N_{i-1}\sigma_i f_i - N_i \left(\sigma_i f_i+\frac{1}{\tau_i}\right), 
\label{kinet}
\end{equation}
where $\sigma_i$ is the cross section for absorption of the $i$th photon, $f_i$ is the photon flux in units of photons cm$^{-2}$~s$^{-1}$, and $\tau_i$ is the characteristic time for the level to depopulate as a result of spontaneous emission and non-radiative relaxation. In the case of a short laser pulse with high intensity, the populations of the excited levels are low ($N_3<<N_2<<N_1<<N_0$), and we can retain only the first term on the right-hand side of the kinetic equations. A number of molecules $N_3$ end up in the dissociative state ($i = 3$), which
Eqs.~\ref{kinet} indicates to be 
\begin{equation}
N_3 = \frac{1}{6}N_0\sigma_1f_1\sigma_2f_2\sigma_3f_3t^3. 
\end{equation}

In terms of the mass-spectrometer signals, substituting specific values for the photon flux and duration
of the laser pulse, we find that the fraction of dissociated molecules is
\begin{equation}
\left(\frac{S}{S_{0}}\right)_{\rm l}=\frac{1}{6}\sigma_1(E_{\rm l}) \sigma_2(E_{\rm l}) \sigma_3(E_{\rm l}) f_{\rm l}^3 \Delta t_{\rm l}^3, 
\label{3photon}
\end{equation}
where $\sigma_1(E_{\rm l})$, $\sigma_2(E_{\rm l})$ and $\sigma_3(E_{\rm l})$ are the cross sections for absorption of the first, second, and third laser photons, and $E_{\rm l} = 5$~eV is the energy of a
laser photon. The photon flux can be found as $f_{\rm l}=F_{\rm l}/E_{\rm l}/\Delta t_{\rm l}$. The cross sections $\sigma_1$ and $\sigma_2$ were measured in \cite{kovacz09} and are equal to $1.4\times10^{-19}$~cm$^{2}$ and $2.8\times10^{-17}$~cm$^{2}$, respectively. The cross section
for stimulated emission $\sigma_3$ can be estimated experimentally from the mass-spectrometer signals for the fragments.

We know from experiment only the fractions of hydrogen atoms $S_{\rm H}$ and acetylene molecules $S_{\rm C_2H_2}$ desorbed from the surface as a result of dissociation. The number of these fragments is less than the total number of dissociated molecules, since dissociation also proceeds along other channels. We can estimate the fraction of these fragments alone for the conditions of the ISM by replacing the signal $S$ in the right-hand side of Eq.~\ref{3photon} with the corresponding
measurements $S_{\rm H}$ and $S_{\rm C_2H_2}$.

In place of the two laser photons required in the experiment for the transition of a molecule into the
quasicontinuum, a single VUV photon would be sufficient. In this case, we are considering only the
values $i = 1, 2$. We assume for $i = 1$ in the kinetic equation (Eq.~$\ref{kinet}$) that $\sigma_1f_1<<1/\tau_1$, and also consider the prolonged action of radiation, when we can move to the stationary case, assuming that the number of molecules excited to state 1 is constant and equal to $N_0\sigma_1f_1\tau_1$. We also neglect stimulated emission in the equation for the dissociative state, as well as deactivation of this state. The fraction of dissociated molecules that come about due to action of the VUV radiation can then be calculated as
\begin{equation}
\left(\frac{S}{S_{0}}\right)_{\rm VUV}=\sigma_{\rm VUV}(E_1) f(E_1) \sigma_{\rm 3}(E_3) f(E_3) \tau_1 \Delta t_{\rm ISRF},  
\label{yield_with_s3}
\end{equation}
where $E_1$ is the energy of the VUV photon, $E_3$ the energy of the photon leading to stimulated emission, $\sigma_{\rm VUV}$ is the cross section for absorption of the VUV photon, $f(E_1)$ and $f(E_3)$ are the number densities of photons with energies $E_1$ and $E_3$, and $\Delta t_{\rm ISRF}$ is the time for the action of the ISRF.\footnote{The index $i = 2$ was replaced by $i = 3$ to bring about agreement in the indices of photons leading to stimulated emission in the presence of UV and VUV radiation.} The cross sections $\sigma_{\rm VUV}$ are presented in \cite{capalbo16}.

In the ISM, the energies $E_1$ and $E_3$ can have any value. To take this into account, we must integrate the ISRF over all possible energies. In this case, Eq.~\ref{yield_with_s3} for the ISRF is transformed as follows:
\begin{equation}
\left(\frac{S}{S_{0}}\right)_{\rm VUV}= \tau_1 \Delta t_{\rm ISRF} \int\int \sigma_{\rm VUV}(E_1) \frac{df_{\rm ISRF}}{dE_1}(E_1) \sigma_{\rm 3}(E_3) \frac{df_{\rm ISRF}}{dE_3}(E_3)G(E_{1}, E_{3}, E_{\rm dis})dE_1,
\label{yield_with_s3_int}
\end{equation}
where  $df_{\rm ISRF}(E_1)/dE_1$ is the spectral flux of photons in photons cm$^{-2}$~eV$^{-1}$, and $G(E_{1}, E_{3}, E_{\rm dis})$ is a function relating the possible energies of the ISRF, the bond dissociation energy $E_{\rm dis}$, and the form of the dissociation curve.

The energy inparted to the molecule, $\Delta E=E_1 - E_3$, must exceed the dissociation energy of the broken bond. For subsequent simplification of our estimates, we assumed that (1) the cross section for stimulated emission $\sigma_{\rm 3}$ does not depend on the photon energy; (2) the energy imparted to the molecule $\Delta E = 5$~eV, as in the experiment; (3) the lifetime of the molecule in the quasicontinuum ($\tau_1$) does not depend on the energies $E_1$ and $E_3$. In reality, it is not known for what value of the potential energy the dissociative state arises; therefore, it is not possible to say at what energy $E_3$ dissociation occurs. However, it follows from the experimental results that the energy $E_1 = 10$~eV corresponds to $E_3 = 5$~eV. We used this difference in energy for any $E_1 \geq 10$~eV. With these assumptions, the unknown function $G(E_{1}, E_{3}, E_{\rm dis})$ takes the form of a delta-function $\delta(E_{1}-E_{3}-\Delta E)$, and Eq.~\ref{yield_with_s3_int} can be rewritten
\begin{equation}
\left(\frac{S}{S_{0}}\right)_{\rm VUV}=  \tau_1 \Delta t_{\rm ISRF}  \sigma_3 \int\limits_{10\;{\rm eV}}^{13.6\;{\rm eV}}  \sigma_{\rm VUV}(E_1) \frac{df_{\rm ISRF}(E_1)}{dE_1} f_{\rm ISRF}(E_1-\Delta E)dE_1,
\label{yield_with_s3_int2}
\end{equation}
where the integration is carried out over energies in the range from 10~eV, corresponding to the minimum energy that is able to lead to dissociation, to 13.6~eV, which is the maximum photon energy in the typical ISM. We will use $k$ to denote the expression on the right-hand side without multiplication signs, which denotes the fraction of molecules that are desorbed per second in the MMP83 field. 

The intensity of the radiation field in a PDR can exceed the mean intensity in the ISM by several orders of magnitude. The intensity at the ionization front in the Orion Bar substantially exceeds the mean value for the ISM, and is equal to $2\times10^4$ in MMP83 units~\cite{goicochea15}, which is a very high value. In the interior of the molecular cloud on whose surface the Orion Bar PDR is observed, this intensity falls practically to zero. We introduced a scaling factor for the radiation intensity $U$ using the MMP83 field as a unit of measurement, and estimated the rate of photodissociation of desorbed molecules in the range from $U = 1$ to $2\times10^5$ (an order of magnitude higher than in the Orion Bar). According to Eq.~\ref{yield_with_s3_int2}, the fraction of desorbed fragments is proportional to $U^2$. Thus, the number of desorbed fragments in the ISM relative to the total number of molecules adsorbed over a time $\Delta t_{\rm ISRF}$ in the presence of a radiation field with intensity $U\times$MMP83 will be
\begin{equation}
\left(\frac{S_{\rm frag}}{S_{0}}\right)^{\rm ISRF}_{\Delta t,U}=  k U^2 \tau_1 \Delta t_{\rm ISRF}. 
\label{frac_isrf_dt}
\end{equation}
The parameter $\tau_1$ in Eq.~\ref{frac_isrf_dt} is the lifetime of the intermediate vibrational level in the quasicontinuum, but the value of this parameter is unknown. All the results presented below were obtained for $\tau_1 = \Delta t_{\rm l}$, but the results can be scaled if this parameter is changed.

\begin{figure}
	\includegraphics[width=0.45\textwidth]{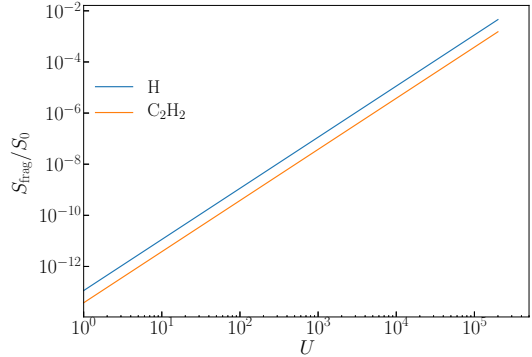}
	\caption{Dependence of the number of desorbed hydrogen atoms and acetylene molecules as a fraction of the total number of adsorbed benzene molecules on the intensity of the interstellar radiation field expressed in terms of the scaling factor $U$.}
	\label{frac_destr}
\end{figure}

Fig.~\ref{frac_destr} shows estimates of the fraction of desorbed hydrogen atoms and acetylene molecules as a function of $U$. It is obvious that, even at the highest radiation intensities, the fraction of desorbed fragments in the ISM remains low, no more than 1\%. In the context of the evolution of dust and its destruction by UV radiation, this means that the effect of the dissociation of adsorbed aromatic molecules is small, although our results show that the surfaces of large grains can be subject to a certain photoerosion. Note that the dependence in Fig.~\ref{frac_destr} and the subsequent
figures is essentially linear, suggesting the possibility of presenting the results as analytical expressions that are suitable for use in astronomical models. However, as was already noted above, given all the uncertainties, we are currently interested only in order-of-magnitude estimates, which are best illustrated by such diagrams.

Let us consider further how important the photodissociation of adsorbed molecules can be in the
context of the chemical evolution of a medium. In particular, acetylene can participate in subsequent
chemical reactions, leading to the formation of observed simple hydrocarbons. Observations of C$_2$H, c-C$_3$H$_2$, and C$_4$H molecules are described in \cite{pety05}, where it is shown that the peak intensities of lines of these molecules coincide with the peak PAH emission. On this basis, it was proposed that the source of these hydrocarbons was the destruction of PAHs. We can see that acetylene could also come about as a result of the dissociation of molecules adsorbed on grain surfaces. We will now compare the output of acetylene due to the destruction of gas-phase PAHs and desorption from grain surfaces as a result of the dissociation adsorbed molecules.

We estimated the absolute number of fragments desorbed from the surfaces of interstellar grains assuming that the entire mantle of a grain consists of benzene molecules. We took the thickness of the
mantle to be half the grain radius, and considered a grain with radius $a = 1$~$\mu$m as an example. To calculate the number of benzene molecules in the mantle, we took the density of hydrogenated amorphous carbon to be 1.4~g~cm$^{-3}$~\cite{jones13} and the molar mass of the molecules to be $\mu=78$~amu. The number of benzene molecules $N_{\rm{C}_6{\rm H}_6}$ in the mantle of a grain with a radius of 1~$\mu$m is then about $4\times10^{10}$. Over a time $\Delta t_{\rm ISRF}$ with $U = 2\times 10^4$~MMP83, about $2\times10^{6}$ hydrogen atoms and $7\times10^{5}$ acetylene molecules are desorbed from this surface. Thus, if this process occurs in the ISM under the specified conditions, the absolute output of fragments will be non-zero.

For comparison, we considered a typical PAH molecule with $N_{\rm C}=50$ carbon atoms. According
to \cite{murga16a}, the rate of production of acetylene $R_{\rm PAH}$ for such a molecule with  $U=2\times10^4$~MMP83 can be roughly estimated as $10^{-7}$~molecules~s$^{-1}$. That is,
such a molecule will be completely destructed in one year. Taking into account the fact that a specific
PAH molecule can produce a number of acetylene molecules that is no higher than half the number of
carbon atoms in the molecule, we find an output of acetylene of about 25 molecules per PAH particle. As was considered above, under the same conditions, the mantle of a grain with a radius of 1~$\mu$m can produce $7\times10^{5}$ acetylene molecules, which is substantially higher.

However, small grains are more abundant than large grains in the ISM. Therefore, it is more correct
to compare the rates of formation of acetylene taking into account the numbers of grains with various
sizes. Let us consider the grain size distribution, $dn/da$, from \cite{wd01}, supposing that all grains with radii smaller than 10~\AA{} are PAHs that can be destroyed in accordance with the model of \cite{murga16a}, and that all the remaining grains are coated in mantles of benzene molecules. The rate of production of acetylene molecules from \cite{murga16a} for PAHs with various sizes is
proportional to $U$. The number of acetylene molecules obtained via the destruction of PAHs in the gas phase is given by

\begin{equation}
N_{\rm C_2H_2}^{\rm PAH} = \Delta t_{\rm ISRF} \int\limits_{a_{\rm min}}^{10\mbox{\,\scriptsize \AA}} R_{\rm PAH} \frac{dn}{da}da,
\label{n_acet_pah}
\end{equation} 
where $a$ is the grain radius, and $a_{\rm min}$ is the minimum grain radius adopted in the distribution~\cite{wd01}.

The number of acetylene molecules desorbed from surfaces of dust grains due to the dissociation of
benzene can be calculated using the formula 
\begin{equation}
N_{\rm C_2H_2}^{\rm mantle} = \int\limits^{a_{\rm max}}_{10\mbox{\,\scriptsize \AA}} \left(\frac{S_{\rm frag}}{S_{0}}\right)^{\rm ISRF}_{\Delta t,U} N_{{\rm C}_6{\rm H}_6}(a)\frac{dn}{da}da,
\label{n_acet_mantle}
\end{equation} 
where $a_{\rm max}$ is the maximum grain size adopted in \cite{wd01}. 

\begin{figure}
	\includegraphics[width=0.45\textwidth]{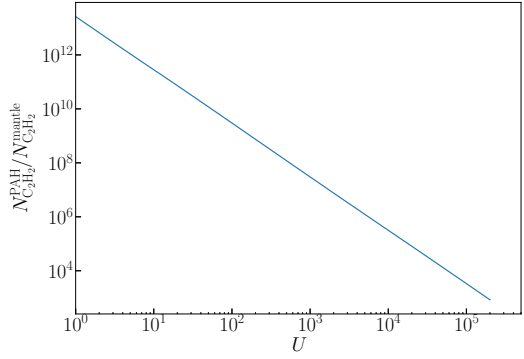}
	\caption{Dependence of the ratio of the number of acetylene molecules produced via the destruction of PAHs in the gas phase to the number of acetylene molecules desorbed from grain surfaces via the dissociation of benzene in grain mantles on the scaling factor $U$.}
	\label{ratio_acet}
\end{figure}

Figure~\ref{ratio_acet} shows that, taking into account the grain size distribution, the number of acetylene molecules produced via the destruction of PAHs in the gas phase is much higher than the number produced via the destruction of mantles containing benzene molecules. Much more acetylene molecules can be produced from the surface of one grain than from the destruction of one PAH particle, but the number of PAH particles is an order of magnitude higher than the number of large grains: for example, the difference in the numbers of grains with radii of 10 and 1000~\AA{} is about eight orders of magnitude. Thus, the dissociation of adsorbed molecules cannot be the main source
of small hydrocarbons in PDRs.

\begin{figure}
	\includegraphics[width=0.45\textwidth]{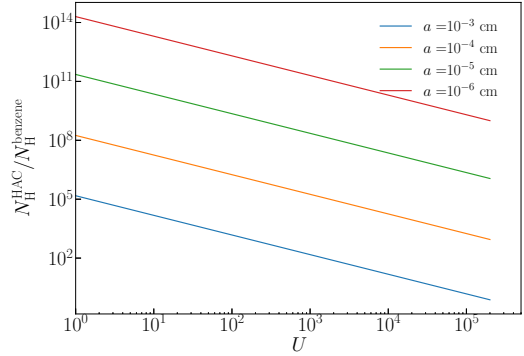}
	\caption{Dependence of the ratio of the number of hydrogen atoms desorbed from a HAC surface to the number desorbed from a benzene mantle on the scaling factor $U$ for various grain sizes.}
	\label{ratio_hac_lab}
\end{figure}

In modern models for the evolution of dust~\cite{jones13, murga16a}, the photodestruction of large grains is treated not in the context of vaporization of the grain mantles, but instead only for grains of hydrogenated amorphous carbon (HAC), from whose surfaces only hydrogen atoms are emitted, so that the grain material gradually becomes poorer in hydrogen. Figure~\ref{ratio_hac_lab} shows the ratio of the number of desorbed hydrogen atoms in the HAC model described in \cite{murga16a} to the number estimated from the results of our current study. It is obvious that hydrogen atoms can be emitted from the surface of HAC much more efficiently than from a benzene mantle. In HAC models, the output of dissociation products is linearly related to the intensity, while the dependence in our current study is quadratic; therefore, this ratio decreases with growth in the radiation intensity, but nevertheless remains high. Only for large grains and very high intensities can these processes have comparable rates.

In conclusion, we note that the laser-desorption experiments described above established that defects
in the crystal structure of the substrate can lead to an amplification of the electromagnetic field acting
on the adsorbed molecules and an increase in the rate of photodissociation of adsorbed molecules. The
surfaces of grains in the ISM most likely have defects, which may be more substantial than those in the
laboratory substrates. As a consequence, we expect that the dissociative effect of the radiation field on the surfaces of cosmic dust grains may also be amplified by several orders of magnitude compared to the case of an ideal surface; this problem requires further study.

\section{CONCLUSION}
We have interpreted the results of laboratory experiments on the laser-induced dissociation of
molecules of simple aromatic hydrocarbons in the context of the conditions in the ISM. We have shown
that the rate of destruction of aromatic grain mantles due to the action of the interstellar radiation field is small. This indicates high photostability of aromatic molecules that have precipidated onto the surfaces of rigid bodies. In the absence of a channel for direct photoexcitation of a repulsive electronic state from the ground state under the conditions of the ISM, the dissociation of adsorbed molecules is much less efficient than the corresponding process in the gas phase. The output of acetylene molecules during the photodestruction of the aromatic mantle of a large grain (with a radius of order 1~$\mu$m) exceeds the acetylene output during the destruction of PAH particles; however, the appreciably higher number density of PAHs compared to that of large grains means that the latter cannot be an important source of acetylene molecules in the gas-phase ISM.

In a broader sense, our results may also indicate lower efficiency than in the gas phase for the photodissociation of other molecules for which there is no direct channel for photodissociation from the ground state. In addition, the products of dissociation in the gas phase and desorption from grain mantles can differ. This means that the use of parameters for gas-phase photoreactions when modeling photoprocesses in dust-grain mantles can potentially lead to appreciably incorrect results.
 
Our results indicate the need for deeper studies of the processes considered, in particular, the destruction of multi-layer adsorbed coatings consisting of both pure molecules and mixtures of molecules that are closest to the compositions of dust-grain mantles in the ISM. As the experiments show, the presence of other molecules in the mantle can give rise to diverse photoprocesses in the mantles~\cite{bouwman11} (without changing the conclusions we have drawn in our current study). 
 
This work was supported by the Russian Science Foundation (project 18-13-00269).

\section*{REFERENCES}
\bibliography{refs_chem}

\end{document}